# Nuclear-spin evidence of insulating and antiferromagnetic state of CuO$_2$ planes in superconducting Pr$_2$Ba$_4$Cu$_7$O$_{15-\delta}$

Sotaro Nishioka[1], Susumu Sasaki[2,3]*, Shunsaku Nakagawa[1], Mitsuharu Yashima[1], Hidekazu Mukuda[1], Mamoru Yogi[4], Jun-ichi Shimoyama[5]

[1]*Graduate School of Engineering Science, Osaka University, Osaka 560-8531, Japan*

[2] *Department of Materials Science, Niigata University, Niigata 950-2181, Japan*

[3]*Japan Agency for Medical Research and Development, Tokyo 100-0004, Japan*

[4]*Faculty of Science, University of the Ryukyus, Okinawa 903-0213, Japan*

[5]*Department of Physical Sciences, Aoyama-Gakuin University, Kanagawa 252-5258, Japan*

E-mail: susumu@eng.niigata-u.ac.jp

In contrast to the "Pr-issue" that neither PrBa$_2$Cu$_3$O$_7$ (Pr123) nor PrBa$_2$Cu$_4$O$_8$ (Pr124) shows superconductivity (SC), we have observed 100%-fraction of SC in an oxygen-reduced Pr$_2$Ba$_4$Cu$_7$O$_{15-\delta}$ (Pr247) which has a hybrid structure that Pr247 = Pr123 + Pr124. It is found that Cu nuclear-spin signals from the CuO$_2$ planes observed at 300K are completely wiped out down at 2K. Instead, plane signals at 2K are observed at higher frequencies. This indicates that, despite the bulk SC, the CuO$_2$ planes in Pr247 are found to be in an insulating and antiferromagnetically ordered state.

Although the mechanisms of "high-$T_c$" cuprate superconductors still remain to be seen, it is widely believed that the $CuO_2$ planes drive the superconductivity (SC).[1] This is because there exist cuprate superconductors that have only $CuO_2$ planes without Cu-O chains. Another well-known fact is the so-called "Pr issue". In both $R$Ba$_2$Cu$_3$O$_7$ (Pr123) and $R$Ba$_2$Cu$_4$O$_8$ (Pr124), where $R$ represents rare earth, SC is observed even in the case that $R$ is magnetic such as Gd, with the only exception for $R$=Pr materials.[1] So far, there have been some reports of SC for Pr$_2$Ba$_4$Cu$_7$O$_{15-\delta}$ (Pr247) that consists of both Pr123 and Pr124 (i.e., Pr247 = Pr123 + Pr124).[2, 3] However, the superconducting fractions of Pr247 reported so far were no greater than 10%.

Recently, we have succeeded in preparing superconducting Pr247 with the shielding fraction of approximately 100%, with the onset SC transition temperature $T_c$ of 18 K. For Pr247, it has been suggested that the SC is driven not by the planes but the Cu-O double chains.[2, 3] In this case, it is expected that, even below the $T_c$, the planes are in an insulating state while the chains are in a superconducting state. As is well known, nuclear-spin measurements play a great role to clarify this, because the signals of nuclear spins at different sites are observed in different resonance frequencies.[1] Moreover, for cuprates, it has been well-established that the Cu resonance frequencies are dominantly given by the configuration of neighboring atoms.[4] Thus, the Cu frequencies observed for Pr247 are likely to be nearly the same as those in $R$123 or $R$124 ($R$=Y, Pr).

In this Letter, we report that the Cu nuclear-spin signals, coming from the $CuO_2$ planes in Pr247 observed at 300K, are completely wiped out down at low temperature of 2K. Instead, the plane Cu signals at 2K are observed at higher frequencies of 60-130 MHz. This indicates that the $CuO_2$ planes, which are paramagnetic at 300K, are antiferromagnetically (AF) ordered at 2K. Taking into account that the internal magnetic field of the AF ordered state is found to be as large as that in the insulating cuprates,[5] it is natural to argue that the AF ordered Cu planes are insulating.[5, 6] Therefore, we have clarified that the $CuO_2$ planes in Pr247 are in an insulating and AF state, despite the emergence of bulk SC with the faction of 100%.

A polycrystalline sample of Pr247 was prepared by calcination of the starting materials Pr$_6$O$_{11}$, BaO$_2$, and CuO at 850℃ for 16 hours in flowing O$_2$ (0.1 %) / Ar gas. It was sintered for phase formation of Pr247 at 950 ℃ for 30 hours in a quartz ampoule, and then it was quenched to room temperature. Finally, the sample was reductively annealed at 520 ℃ for 16 hours in a flowing Ar gas and at 700 ℃ for 175 hours in an evacuated quartz ampoule. Then, it was quenched to room temperature again. It was found to be of single phase from X-ray diffraction. The oxygen reduction amount $\delta$ was estimated to be 0.65 from thermo gravimetric analysis. As shown in Fig. 2, the static magnetic susceptibility $\chi$ at a magnetic field of 1 Oe starts to show a superconducting transition below 18 K. The magnetic susceptibility ($4\pi M/H$), in the process of zero-field cooling, decreased to -1.0, suggesting the shielding volume fraction of almost 100 %, substantially larger than those in the previous studies.[2, 3]

For the Pr247 with a 100%-fraction of SC, we performed Cu nuclear-spin measurements at zero external field, i.e., nuclear quadrupole resonance (NQR). To obtain Cu-NQR frequency spectra, we have taken the signal intensities point-by-point with varying frequency. This is because, unlike the case of liquid, the linewidths in solids are generally larger than the frequency range that a single pulse can fully irradiate. For the point-by-point measurements, we built ourselves an auto-tuning system that controls both the resonance and matching capacitors in the tank circuit of the probe. To simulate the spectra in the presence of multiple parameters, we employed an evolutionary algorithm based on a Covariance Matrix Adaptation Evolution Strategy (CMA-ES), which is a kind of artificial intelligence (AI) that gives the optimum solution, avoiding artifacts coming from "local minima" in the process of the standard least-square methods.

Before discussing the Cu-NQR spectra, we briefly comment on the Cu sites of Pr247. Since Pr247= Pr123 +Pr124, there exists four crystallographically inequivalent Cu sites. As illustrated in Fig. 1, the Pr123 unit has a plane site and a single chain site, and the Pr124 unit has a plane site and a double chain site. We define the site in the single chain as Cu(1), the plane site in Pr123 (Pr124) unit as Cu(2) [Cu(3)], and the double chain sites as Cu(4). Figure 3 shows the Cu-NQR spectra for Pr247 at 300 K and 2 K. In the absence of magnetic field, the peak frequency, $\nu_Q$, is generally given as $\nu_Q = eQV_{ZZ}/2h$, where $eV_{ZZ}$ is the electric field gradient at the nucleus, $Q$ is the quadrupolar moment of the nucleus, and $h$ is the Planck constant. In the case of cuprates, it is now well established that the configuration of neighboring atoms around a Cu site primarily gives the electric field gradient $eV_{ZZ}$, and hence $\nu_Q$.[4] Thus, we can safely assume that the peak frequencies of Fig. 3 are nearly the same as those in $R$123 or $R$124 cuprates previously reported.[4, 7] For the assignments, we can utilize the universal fact on isotopes: the $^{63}$Cu ($^{65}$Cu) nucleus has a spin of $I = 3/2$, an electric quadrupole moment $^{63}Q$ = -0.211 ($^{65}Q$ =-0.195) $\times$ $10^{-30}$ m$^2$, a gyromagnetic ratio of $^{63}\gamma_n$ = 11.285 ($^{65}\gamma_n$ =12.089) MHz/T, and a natural abundance of 69.1 % (30.9 %). Thus, the ratio of $^{63}\nu_Q/^{65}\nu_Q$ is exactly the same as $^{63}Q/^{65}Q$, and the intensity ratio is the same as the ratio of the natural abundance of 69.1/30.9.

As indicated in Fig. 3, the sharp peak at 20.7 MHz (19.1 MHz) is assigned to $^{63}$Cu(4) [$^{65}$Cu(4)] sites in the double chain.[7,8] The other sharp peak at 30.2 MHz (27.9 MHz) is assigned to $^{63}$Cu(1) [$^{65}$Cu(1)] in the single chain.[4] Here, it should be noted that the single- or double-chain signals at 300 K are observed at nearly the same frequencies down at 2K. This indicates that the local electronic states at single- and double-chain sites are non-magnetic, because in the presence of magnetic field, the peak frequencies should depend on not only the $\nu_Q$, but also the magnetic field.

The large sharp peaks at 22 MHz and 31.7 MHz observed at 300K are assigned to the plane sites $^{63}$Cu(2) in the Pr123 unit and $^{63}$Cu(3) in the Pr124 unit, respectively. This is due to the fact that those frequencies are nearly the same as those in $R$123(6) [4, 9] and $R$124 [7, 8]. Here it should be noted that, in contrast to the Cu chain site, these Cu signals from the planes observed at 300K are completely wiped out down at 2 K, as seen in Fig.3. Instead, signals appear in the higher frequencies of 60-130 MHz, where no signals are observed at 300K. The broad signals at higher frequencies indicate the appearance of the internal magnetic field $B_{int}$. In general, the Hamiltonian for Cu nuclear spins ($I = 3/2$), in crystal lattices with an axial symmetry, is described by the Zeeman term $H_Z$ and the nuclear-quadrupole interaction term $H_Q$ as

$$H = H_Z + H_Q = -\gamma_n \hbar \mathbf{I} \cdot \mathbf{B} + h\nu_Q/6 \cdot [3I_z^2 - I(I+1)].$$

With this formula, the higher-frequency spectrum at 2 K can be reproduced by the superposition of the spectrum from two Cu sites that overlap at the wide frequency range, as shown by red solid curves in Fig.3. Here, the parameters for the Cu(2) site are found to be that the internal magnetic field $B_{int}(2)$= 9.36 T, the angle of $B_{int}(2)$ from c-axis $\theta_B(2)$= 82.0 degree, FWHM (2) = 4.86 MHz (blue bars on the frequency axis at 95-130 MHz). For the Cu(3) site, $B_{int}(3)$= 7.11 T, $\theta_B(3)$= 78.1 degree, FWHM (3) =5.09 MHz (orange bars on the frequency axis at 60-100 MHz). Here we would like to emphasize that $^{63}\nu_Q(2)$=22 MHz and $^{63}\nu_Q(3)$=31.7 MHz are not fitting parameters, but given solely from the spectrum at 300K. Thus, we can safely conclude that both CuO$_2$ plane sites, Cu(2) and Cu(3), are AF ordered although Pr247 exhibits bulk SC of the fraction of 100%.

It may be tempting to argue that, even in the presence of AF order, the $CuO_2$ planes can be metallic.[6] However, this is not the case, because the AF ordered signals of Pr247 are observed at frequencies as high as those in the insulating states of cuprates.[4, 5] If the planes of Pr247 were metallic in the presence of AF order, the plane Cu signals should appear at lower frequencies as in multilayered cuprates.[6] For more clarity, we discuss it quantitatively: the $B_{int}$ is proportional to the AF moment $M_{AF}$, as given by $B_{int} = |A_{ab} - 4B| M_{AF}$, where $A_{ab}$ is the onsite hyperfine field and $B$ is the super-transferred hyperfine field from four nearest-neighbor Cu sites. If we tentatively assume the values of $A_{ab} = 3.4$ T/$\mu_B$ and $B = 4.1$ T/$\mu_B$ estimated for Y123(6.63)[10], we obtain the $M_{AF}$ value of 0.72 $\mu_B$ for Cu(2) in the Pr123 unit, and 0.55 $\mu_B$ for Cu(3) in the Pr124 unit. The values of $B_{int}$ or $M_{AF}$ are nearly the same as those of the insulating AF (or Mott) state of Y123(6)[4, 5, 11] and Pr124.[7] Therefore, we can safely conclude that the $CuO_2$ planes in the superconducting Pr247 are in an insulating and AF ordered state.

Next we discuss the oxygen deficiency $\delta$ (~0.65) in Pr247. In the case of $YBa_2Cu_3O_{7-\delta}$ (Y123(7-$\delta$)), it is known that the oxygen deficiency in the single chain Cu(1)-$O_{1-\delta}$ significantly alters the ground state of $CuO_2$ plane: from the high-$T_c$ state with $T_c$ ~90K for $\delta$ ~0 to the insulating AF (or Mott) state for $\delta$ ~1. In the case of $RBa_2Cu_4O_8$ ($R$124), the oxygen at double $Cu(4)_2O_2$ chains are generally known to be crystallographically robust, and hence no oxygen deficiency is expected in the $R$124 unit of Pr247. Thus, in the Pr247 structure, the oxygens are likely to be reduced from the single chains in the Pr123 unit. As a result, in the single chains, in addition to the Cu(1) sites, other two Cu sites can be possible. One is Cu(1') site that has one oxygen vacancy at either side along the chain, as in O-Cu(1')-(vacancy). For the other possible Cu(1") site, both of the neighboring oxygen sites are occupied, as in O-Cu(1")-O. However, the Cu(1") can be observed only in the case that the oxygen occupancy is larger than 0.65 (i.e., $\delta$ <0.35) in Y123(7-$\delta$)[4], which is consistent with the definition of Cu(1"). In contrast, for the present Pr247(15-$\delta$), the oxygen occupancy is 0.35 ($\delta$ ~0.65), which is substantially smaller than 0.65 where Cu(1") signals can be observed.[4] Accordingly, we can safely say that the single chain Cu site in Pr247(15-$\delta$) is dominated by Cu(1), with a smaller probability of Cu(1') and no Cu(1"). Therefore, the smaller and broader signals of 23 MHz are naturally assigned to $^{63}Cu(1')$. As a result, the signal from the isotope $^{65}Cu(1')$, which is expected to be half intensity of $^{63}Cu(1')$ at 21.3 MHz, is naturally considered to be merged with the tail of $^{63}Cu(4)$ signal.

Finally, we discuss where in the Pr247 drives the SC. As already clarified, unlike all the other cuprate superconductors, the CuO$_2$ planes of Pr247 are in an insulating and AF ordered state. Here it is to be reminded that the double chain site is metallic even in non-SC Pr124,[12] Pr248, Pr247(15-δ),[13] Pr123 and Pr1248,[14] despite the fact that the CuO$_2$ planes are in an insulating AF state as in Pr247. Thus, it is very likely that the double chains drive the SC of Pr247. Since neither Pr123 nor Pr124 exhibit SC, we have a picture that, for the emergence of SC in Pr247, the carriers are doped into the double chains Cu(4)$_2$O$_2$ as a result of the oxygen reduction from the single chain Cu(1)-O.

In conclusion, despite the fact that the Pr-substituted cuprate of Pr123 or Pr124 does not exhibit SC, we have successfully synthesized a single-phase superconductor Pr247 with the shielding fraction of 100% and the onset $T_c$ of 18 K. We have found that Cu plane signals observed at 300K are completely wiped out down at 2K. Instead, Cu plane signals are observed at higher frequencies of 60-130 MHz. Since the spectrum at 2K is well reproduced by the Mott insulating picture, we can safely say that, despite the bulk SC of Pr247, the planes are in an insulating and AF state.

This research was performed mainly by JSPS KAKENHI Grant 19H02580 (S. S) and 18K18734 (H. M). H. M. is also supported by the Mitsubishi Foundation and the Tanigawa Fund. S.S is supported also partly by the joint research program of Research Institute of Electrical Communication (RIEC) Tohoku University, the collaboration program of Institute of Materials and Systems for Sustainability (IMaSS) from Nagoya University, the Naito Scholarship Foundation and the Yamaguchi Educational and Scholarship Foundation.

## Figure Captions

**Fig. 1.** Structure of $Pr_2Ba_4Cu_7O_{15-\delta}$ (Pr247). Pr247 consists of the alternative stacking of Pr123 and Pr124 unit. Pr247 has four inequivalent Cu sites, which are the single chain Cu(1) site, the plane site Cu(2) in the Pr123, the plane site Cu(3) site in the Pr124 unit, and the double chain Cu(4) sites. The Cu atoms in the double chain do not form a "ladder" structure but a "double" chain or "zigzag" chain.

**Fig. 2.** Temperature dependence of magnetic susceptibility of Pr247 under zero-field cooling (ZFC) and field cooling (FC) at a magnetic field of 1 Oe. The diamagnetic response of SC transition appears gradually below 18 K (the onset $T_c$), resulting in the shielding fraction of 100% well below the $T_c$.

**Fig. 3.** Cu-NQR spectra for superconducting Pr247 at 300 K (top) and 2 K (bottom). The peaks are assigned to inequivalent Cu(i) (i=1-4) sites based on the results of bulk $R$123 and $R$124.[4, 7, 8, 9] The signals from the planes, Cu(2) in the Pr123 unit and Cu(3) in the Pr124 unit at 300K, are completely wiped out at 2K. Instead, signals appear at 60-130MHz, as a result of AF order in the planes. In contrast to the planes, no changes are observed in the chain frequencies: Cu(1) in the single chain and Cu(4) in the double chain. The red solid line is the best-fit simulation of the spectra in an AF ordered state (see text). The smaller and broader signals at around 23 MHz down at 2K can be attributed to Cu(1') site, where either side of the neighboring oxygens sites is vacant, as a result of oxygen reduction from the single chains.

Fig. 1

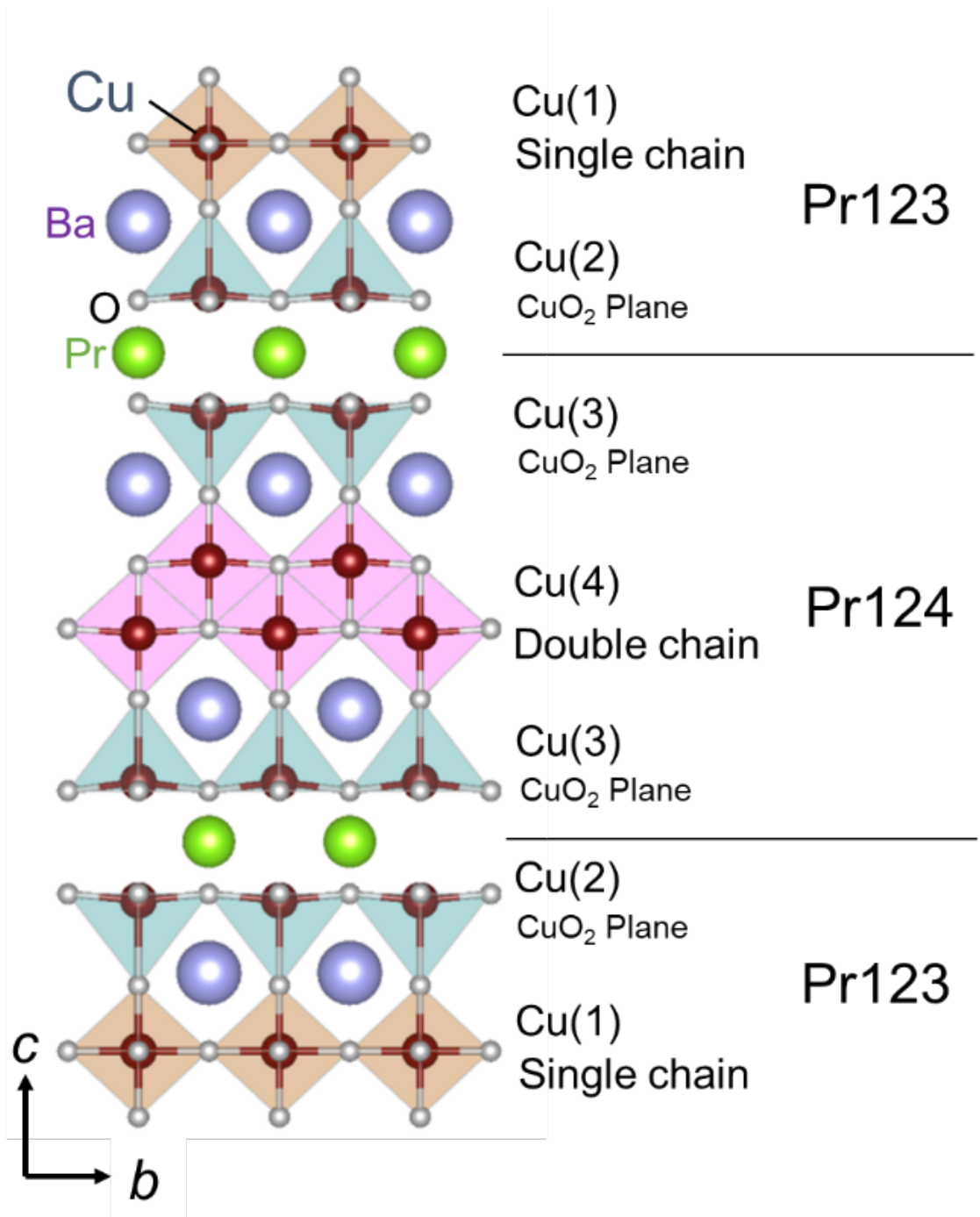

Fig. 2

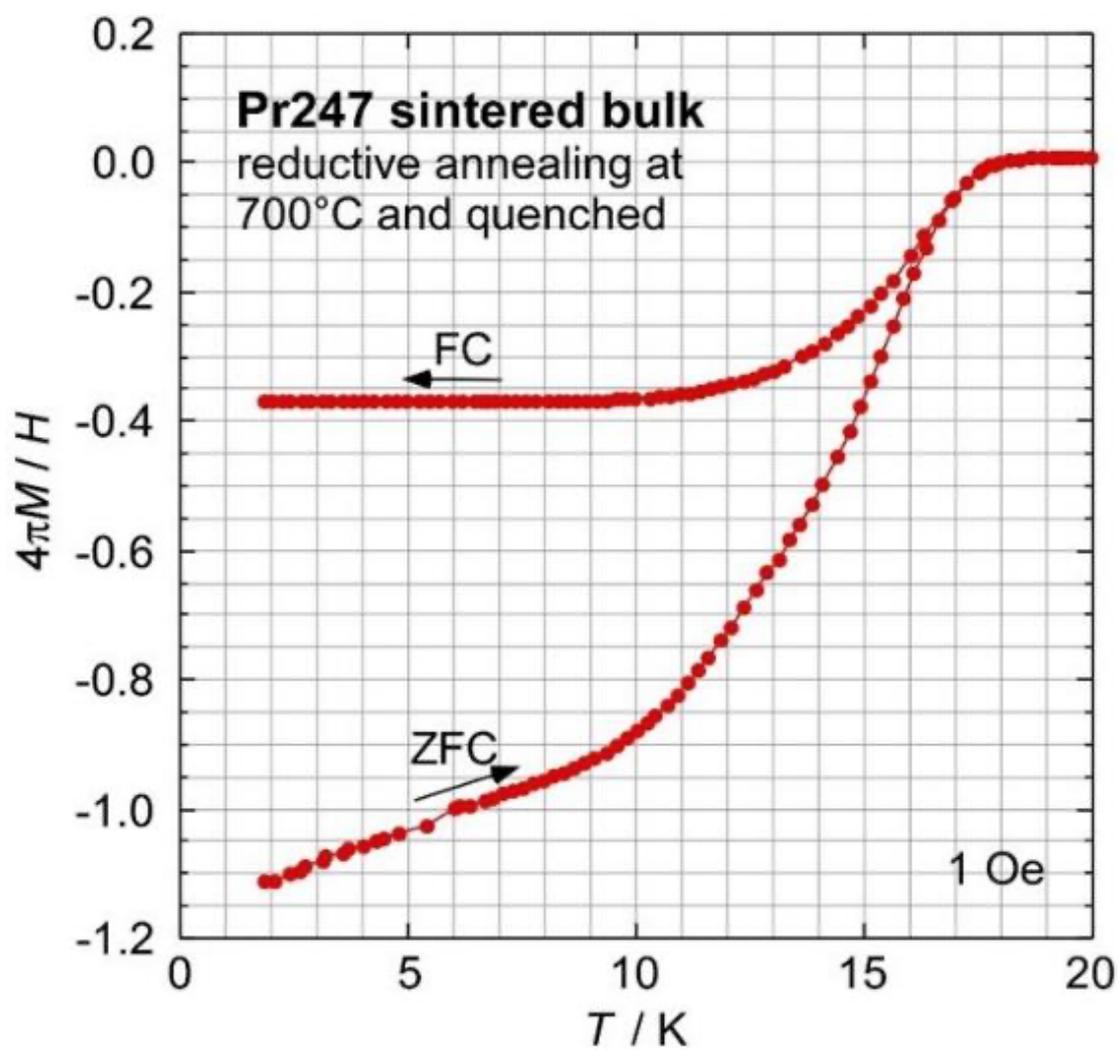

Fig. 3

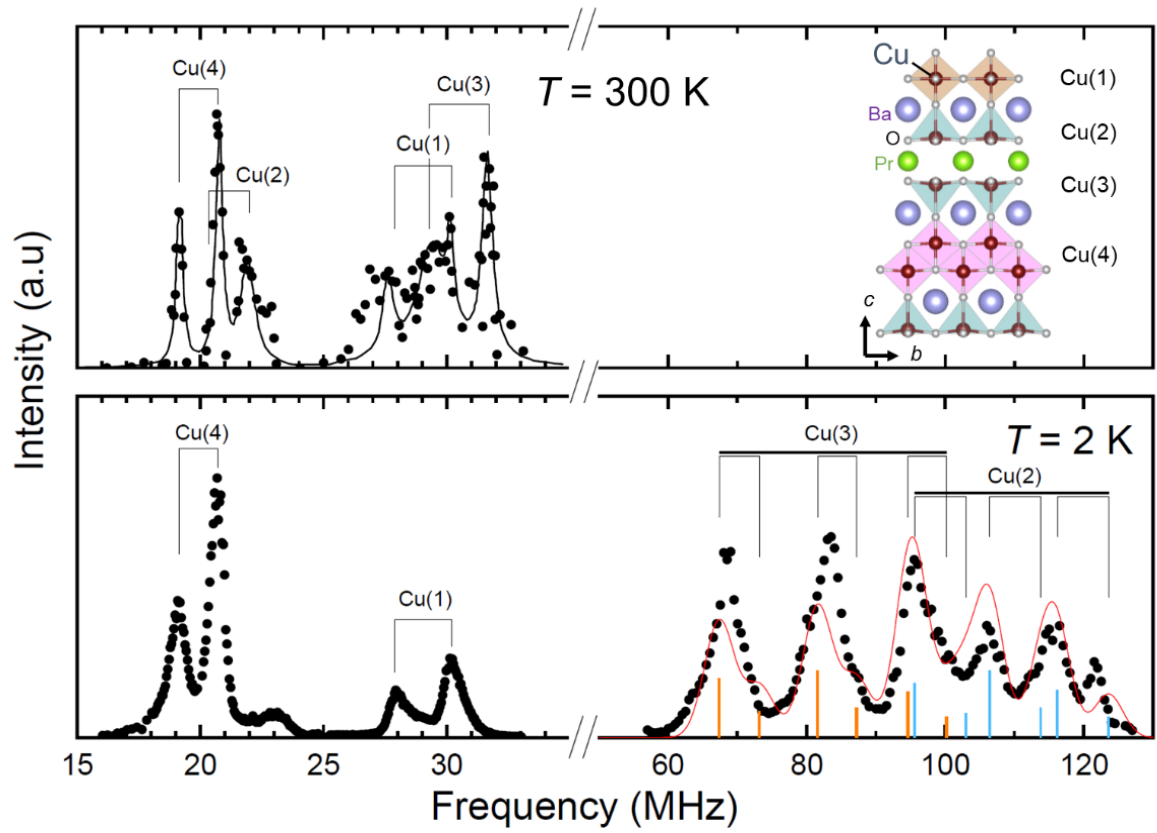

**Supplementary Information**

From the XRD pattern shown below, the present Pr247 is of single phase, without any extrinsic phases. This indicates that the superconductivity (SC) of Pr247 is intrinsic, excluding the possibility that extrinsic phases cause SC.

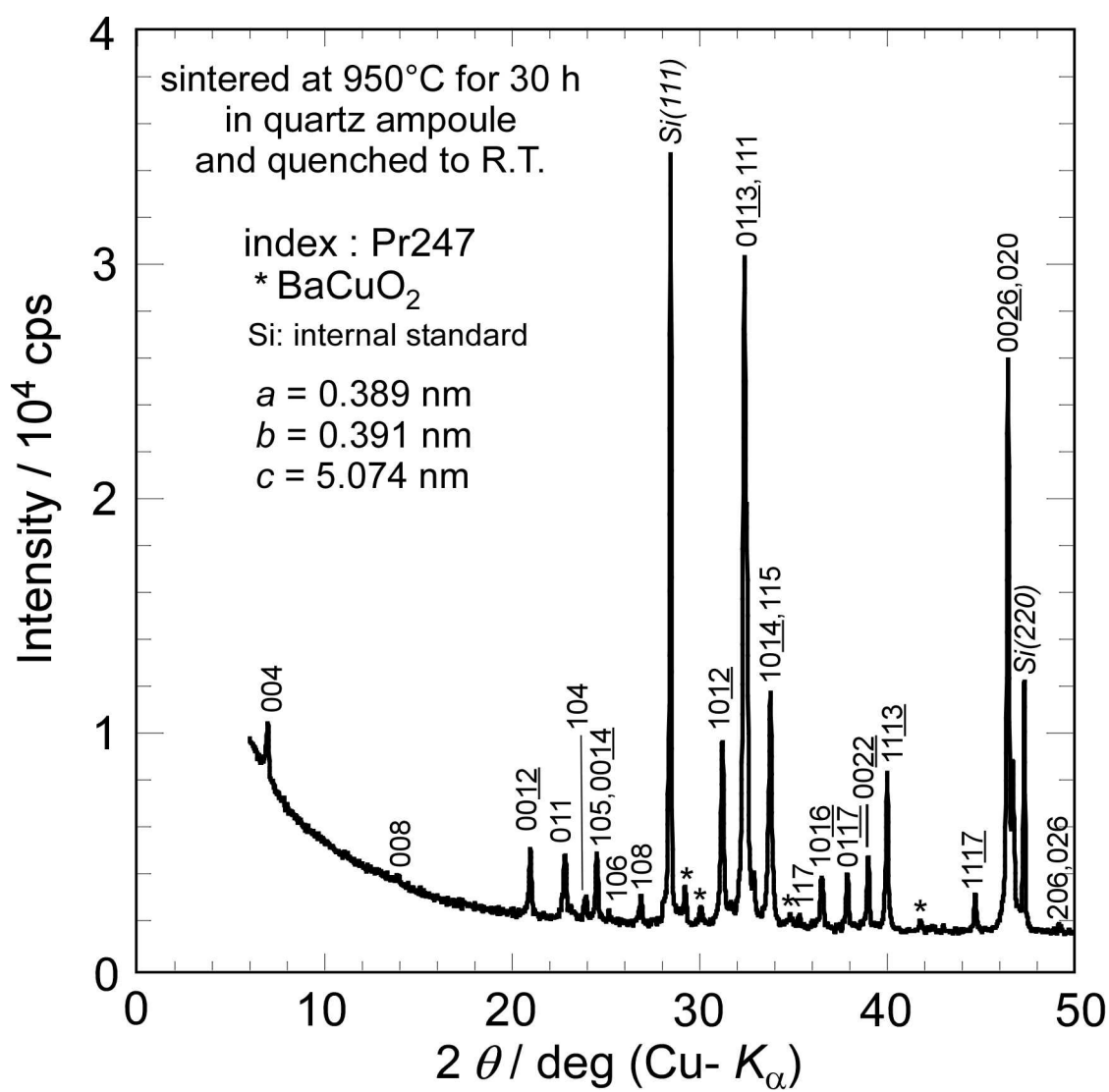